\documentclass[%
 aip,jap,
 amsmath,amssymb,
reprint,%
]{revtex4-1}

\usepackage{graphicx}% Include figure files
\usepackage{dcolumn}% Align table columns on decimal point
\usepackage{bm}% bold math
\begin{document}

%\preprint{AIP/123-QED}

\title{Superconducting transition edge sensors with phononic thermal isolation}

\author{E.A. Williams}
\author{S. Withington}
\author{C.N. Thomas}
\author{D.J. Goldie}
\affiliation {Cavendish Laboratory, University of Cambridge, J.J. Thomson Avenue, Cambridge CB3 OHE, United Kingdom}
\author{D. Osman}
\affiliation{Now at Sention, Pyramid House, 954 High Road, London N12 9RT, United Kingdom}

\date{\today}

\begin{abstract}
The sensitivity of a low-noise superconducting transition edge sensor (TES) is determined by the thermal
conductance of the support structure that connects the active elements of the device to the heat bath.
Low-noise devices require conductances in the range 0.1 to 10\,pW\,K$^{-1}$, and so have to rely on diffusive
phonon scattering in long, narrow, amorphous  SiN$_{\rm x}$ legs. We show that it  is possible to
manufacture and operate TESs having short, ballistic low-dimensional legs (cross section 500 $\times$ 200\,nm)
that contain multi-element phononic interferometers and ring resonators. These legs transport heat in effectively 
just 5 elastic modes at the TES's operating temperature ($<$ 150\,mK), which is close to the quantised limit of 4. 
The phononic filters then  reduce the thermal flux further by frequency-domain filtering. For example, a micromachined
3-element ring resonator reduced the flux to 19 \% of a straight-legged ballistic device operating at the
quantised limit, and 38 \% of a straight-legged diffusive reference device. This work opens the way to
manufacturing TESs where performance is determined entirely by filtered, few-mode, ballistic thermal transport
in short, low-heat capacity legs, free from the artifacts of two level systems.
\end{abstract}

\maketitle

\section{Introduction}

There is considerable interest in developing superconducting Transition Edge Sensors
(TESs)\cite{IrwinChapter} for astronomy and space science. For ground-based photometric
measurements at long wavelengths (3 mm - 300\,$\mu$m), Noise Equivalent Powers (NEPs) of
10$^{-17}$\,WHz$^{-1/2}$ are  required \cite{posada2015fabrication,westbrook2016development,AlMnAdvancedACTPol,BICEP2KECKSPIDER}; for space-based measurements and Earth Observation
at long wavelengths, NEPs of 10$^{-18}$\,WHz$^{-1/2}$ are necessary; for space-based
measurements with cooled-aperture telescopes at FIR wavelengths, such as SPICA (200 - 30\,$\mu$m)
\cite{SafariSPICA,SafariNewImproved,SPICANewFramework,goldie2016performance,goldie2012ultra}, NEPs of 10$^{-19}$\,WHz$^{-1/2}$ and better are the necessary target.  Time and energy
resolved photon counting TESs are being developed for the x-ray space telescope Athena (0.2 - 12\,keV)
\cite{gottardi2016development,den2014requirements,akamatsu2016tes,TESMicrocalorimeterATHENA}, and for general utilitarian applications at optical wavelengths (1550 - 400\,nm)
\cite{cabrera1998SinglePhoton,portesi2008OpticalPhoton,eisaman2011invited,SPDNatPhot2009,QKDNoiseFree}.

State-of-the-art TESs have many favourable characteristics, but they also have a number of
shortcomings. To achieve low-noise operation, a low thermal conductance
($G =$ 0.1 - 10\,pW\,K$^{-1}$) is needed between the active elements of the device and the
heat bath. TESs are usually fabricated on SiN$_{\rm x}$ membranes, and thin
($H =$ 200\,nm - 1\,$\mu$m), narrow ($W =$ 1 - 10\,$\mu$m), long ($L =$100 - 700\,$\mu$m) legs
patterned into the membrane, using Deep Reactive Ion Etching (DRIE), to achieve the necessary
thermal isolation. The lower the target NEP, the lower the thermal conductance required, and this
leads to quite extreme geometries. In the case of ultra-low-noise imaging arrays, long legs
($L =$ 500\,$\mu$m - 1\,mm) prevent tight optical packing, and inefficient optical coupling
schemes must be used to minimise the effects of the large pixel-to-pixel spacing. In addition,
SiN$_{\rm x}$ is a highly disordered dielectric and contains an abundance of Two Level Systems
(TLSs) \cite{anderson1972anomalous,phillips1972tunneling,zink2004specific}. TLSs result in a specific heat that is many hundreds of times higher than
the Debye value and, when combined with low thermal conductance, this leads to devices that
are too slow for some applications. Also, phonon trapping in long, narrow legs causes localised
transport, which results in wide variations (at least $\pm$ 15~\%) in the performance of even
notionally identical devices on the same wafer.

In a previous paper we demonstrated that it is possible to manufacture SiN$_{\rm x}$ TESs
having tiny ballistic support legs ($H =$ 200\,nm, $W <  1\,\mu $m, $L = $ 1 - 4\,$\mu$m) \cite{DJBallistic}.
The thermal conductance and thermal fluctuation noise in these devices was found to be fully
predicted by heat transport calculations based solely on the dispersion curves of elastic modes
calculated using the bulk elastic constants of the material. Moreover, the uniformity in performance
was high as a consequence of having eliminated resonant phonon scattering in the disorder of the
material.

At low temperatures ($\leq$ 150\,mK), heat is transported in low-dimensional dielectric bars through
a small number of elastic modes. In our ballistic devices \cite{DJBallistic}, approximately 6-7 modes were excited,
which is close to the quantised limit of 4: one compressional, one torsional, and in-plane and
out-of-plane flexure. In a subsequent series of experiments \cite{ThermalAttenuation2017}, we measured the thermal
elastic attenuation length of these modes to be 20\,$\mu$m, and so our short-legged TESs were operating
well within the ballistic limit. It can be shown, and was found in practice, that the ballistic, few
mode limit corresponds to an NEP of approximately 10$^{-18}$\,WHz$^{-1/2}$. This NEP cannot be reduced
further by increasing the length, because there is no scattering, or reducing the cross section,
because we have already reached the quantised limit. The question arises as to whether it is possible
to incorporate micromachined phononic filters into the low-dimensional legs of low-noise TESs in order
to reduce the NEP below the ballistic quantised limit.

The incorporation of phononic filters would have a number of benefits: First, it should be possible to
manufacture low-$G$ devices having legs that are significantly shorter than their long-legged diffusive
counterparts. Second, the reduction in $G$ would be brought about by a phase coherent scattering
process, which is likely to have a beneficial effect on the thermal fluctuation noise in the legs, as
compared with that generated by a dissipative diffusive process. Third, we would like to manufacture
devices using crystalline Si membranes \cite{ThermalSilicon,SiliconCMB}, as this would significantly reduce the heat
capacity of the device, but the dispersion curves of Si are very similar to those of SiN$_{\rm x}$,
and the elastic attenuation length considerably larger due to the low density of TLSs. Therefore
phononic filters are needed if crystalline Si-membrane devices, which would have exceedingly
long phonon mean free paths, are to be produced having NEPs of better than 10$^{-18}$\,WHz$^{-1/2}$.

The objectives of the exploratory work described here were as follows: (i) to determine whether TESs
having low-dimensional phononic filters can be manufactured at all; (ii) to develop and compare
manufacturing techniques using optical lithography (OL) and electron beam lithography (EBL); (iii) to
investigate whether TESs with phononic filters behave in a conventional way; (iv) to determine whether
thermal conductance can in practice be reduced significantly below the few-mode quantised limit; and (v)
to investigate uniformity in performance between notionally identical devices. The experimental work was
based solely on SiN$_{\rm x}$ membranes, but the results give direct information about the likely
behaviour of phononic devices based on crystalline Si membranes.

\section{Theory}

\subsection{Elastic waves and ballistic thermal power}
\label{subsec:Theory_Elastic}

The thermal flux through a uniform, low-dimensional, ballistic, dielectric bar can be calculated directly
from the dispersion curves of the discrete elastic modes \cite{DJBallistic}. Here we summarise the calculation because it is
central to the subject matter of the paper, and because the ballistic limit will be used later for normalising
experimental data.

The classical elastic wave equation is
\begin{equation}
\rho\omega^2 u_i +C_{ijkl}\frac{\partial^2 u_k}{\partial x_j \partial x_l} = 0,
\label{eq:ElasticWave}
\end{equation}
where $u_i$ is the displacement field in Cartesian direction $i$, $C_{ijkl}$ the fourth-rank
stiffness tensor, $\rho$ the mass density, $\omega$ the angular frequency, and the Einstein
summation notation has been assumed. Equation
(\ref{eq:ElasticWave}) can be solved by adopting a general basis for the displacement field,
\begin{equation}
\label{eq:BasisExp}
u_i =  a_{ir}\psi_{ir},
\end{equation}
where $a_{ir}$ is the $r$'th expansion coefficient of the $i$-directed displacement and $\psi_{ir}$ is
the associated basis function. Equation (\ref{eq:BasisExp}) may be substituted into
Eq. (\ref{eq:ElasticWave}), and the resulting algebraic equations solved numerically to give the dispersion
curves of the propagating modes. Although a variety of basis functions, such as Gaussian-Hermite polynomials,
could be used for this purpose, we have found power-series expansions to be particularly effective \cite{nishiguchi1997acoustic}.

In the case of a homogeneous, isotropic, insulating dielectric such as SiN$_{\rm x}$, the stiffness tensor
simplifies, and the modal calculation requires only the mass density, $\rho$, Young's modulus, $E$,
and Poisson's ratio, $\nu_p$, of the material, in addition to the height, $H$, and  width, $W$, of the
bar. Averaging over any specific microstructure in favour of the bulk elastic properties is
appropriate given the long dominant phonon wavelengths ($>$1\,$\mu$m) at low temperatures.

\begin{figure}
\centering
cd \includegraphics[width=3.37in]{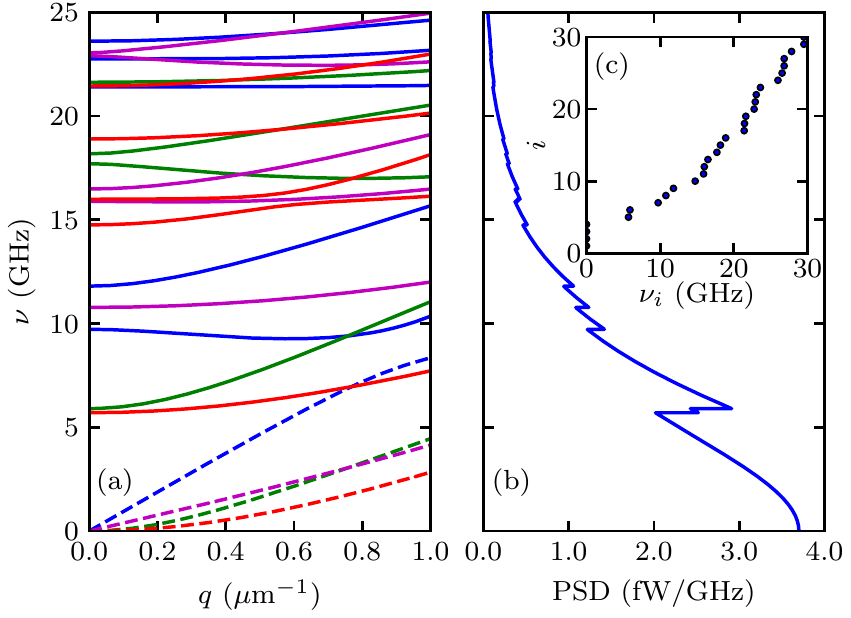}% Here is how to import EPS art
\caption{\label{fig:Modes_Plot} (a) Frequency, $\nu$, against wavenumber, $q$, of elastic modes
with cut-off frequencies up to 25 GHz for a SiN$_{\rm x}$ bar having cross section  500 $\times$ 200\,nm.
Longitudinal, in-plane transverse, out-of-plane transverse and torsional symmetries are shown in blue,
green, red and magenta respectively, with the principal modes dashed. (b) Total net power spectral
density (PSD) for a bar with termination temperatures of 68\,mK and 135\,mK, with the ordinate matching
that of (a). (c) Mode number $i$, against cut off frequency $\nu_{i}$, showing quadratic dependence.}
\end{figure}

Figure \ref{fig:Modes_Plot}(a) shows the dispersion profiles of the low-order modes of the experimentally
considered geometry $H=$\,200 nm, $W=$ 500\,nm, with $\rho =$ 3.14\,g\,cm$^{-3}$, $E=$ 280\,GPa and
$\nu_p =$\,0.28 \cite{vlassak1992new,walmsley2007poisson,MaterialsCRC}. With the exception of the four lowest-order modes, all modes have a cut-off
frequency that increases as the cross-sectional area is reduced. Each mode can be assigned to one of four
2-dimensional displacement symmetries: compressional, in-plane and out-of-plane flexural, and torsional.
The lowest order mode in each group is a principal mode with no cut-off. Propagation of these four principal
modes at all frequencies imposes a fundamental lower limit on the power transmitted ballistically along a
straight bar, even as its dimensions are reduced such that the higher order modes carry negligible power.
This is often called the `quantised limit'.

In the context of TESs, each leg consists of a SiN$_{\rm x}$ bar terminating at the central island
with a temperature taken to be the superconducting transition temperature $T_C$, and at the
surrounding silicon wafer, held at the bath temperature $T_B$. The net thermal power transmitted
from the island to the heat bath in the ballistic limit is therefore obtained by summing over the
power carried by each mode, giving
\begin{equation}
P_{bal} = \displaystyle \sum_i \int_{\nu_i}^\infty B(\nu,T_C) - B(\nu,T_B) \, \mathrm{d}\nu,
\label{eq:Pbal}
\end{equation}
where $\nu_i$ is the cut-off frequency of the $i$th mode and
\begin{equation}
B(\nu,T) = \frac{h\nu}{\mathrm{e}^{h\nu/k_BT}-1}
\end{equation}
is the single-mode Power Spectral Density (PSD). Previously, we have demonstrated a strong agreement
between the net thermal power given by Eq. (\ref{eq:Pbal}) and measurements on TESs with leg lengths
less than 4\,$\mu$m, and a range of widths \cite{DJBallistic}. This work confirmed that the net power
can be calculated from first principles, through the bulk elastic constants, without free parameters,
independently of the precise stoichiometry of the SiN$_{\rm x}$.

Figure \ref{fig:Modes_Plot}(b) shows the total net PSD, $B(\nu,T_C) - B(\nu,T_B)$, summed over all
simulated modes, for $T_C = 135$\,mK and $T_B = 68$\,mK. Sharp discontinuities are evident where individual
modes cut on, corresponding to the intercepts of the dispersion curves with the ordinate of
Fig. \ref{fig:Modes_Plot}(a). At these experimentally representative temperatures, the PSD rolls off
such that modes with $\nu_i \gtrsim 25$\,GHz carry negligible power. Figure \ref{fig:Modes_Plot}(c)
shows mode number $i$ against cut-on frequency $\nu_i$, indicating the number of propagating modes
as a function of frequency. Above the four principal modes, the number of modes increases quadratically
with frequency.

In numerical work, it is convenient to normalise calculated powers to the power carried by a
single principal mode, $P_{qua}$, which defines an effective number of propagating modes:
\begin{equation}
N_{eff} = \frac{P_{bal}}{P_{qua}} = \frac{\displaystyle \sum_i \int_{\nu_i}^\infty B(\nu,T_C) - B(\nu,T_B) \, \mathrm{d}\nu}{\int_{0}^\infty B(\nu,T_C) - B(\nu,T_B) \, \mathrm{d}\nu}.
\end{equation}
$P_{qua}$ is the net power that would be carried by a single elastic mode. In the limit of narrow legs 
operating at low temperatures, the effective number of
modes approaches $N_{eff}=4$. The experimentally measured power $P$ may be substituted for $P_{bal}$, as is 
done in Section \ref{sec:Results_Discussion}, to calculate the actual effective number of modes propagating 
in test structures, $N_{eff}=P/4P_{qua}$, for which the ballistic case is an upper limit. 

In experimental work, it is convenient to normalise the measured
power $P$ flowing from the TES island to the heat bath to the straight-leg ballistic limit,
\begin{equation}
\epsilon = \frac{P}{4N_{eff}P_{qua}}=\frac{P}{4P_{bal}},
\label{eq:Epsilon}
\end{equation}
where the factor of 4 arises because each TES has 4 legs. In the case of a phononic thermal
filter, $\epsilon$ quantifies the level of power attenuation achieved relative to the multi-mode
ballistic case.

\subsection{Phononic interferometers}

The central question of this paper is whether it is possible to achieve a significant reduction
in thermal flux by introducing phononic filters into low-dimensional dielectric bars. In work
on TESs, it is common practice to describe the heat flux in the legs by the equation
\begin{equation}
P = K \left( T_{C}^{n} - T_{B}^{n} \right),
\label{eq:ThermalFlux}
\end{equation}
where $K$ is a parameter that determines the overall magnitude of the flux, and $n$ is a
parameter that describes the functional dependence on temperature. For truly ballistic
transport in a single-mode structure $n=2$, whereas for ballistic transport in a highly
multimode structure $n=4$ \cite{withington2011low}. In general, for diffusive transport in a few-mode
structure, $n$ is intermediate between these two values. It follows from Eq. (\ref{eq:ThermalFlux})
that the differential thermal conductance is given by
\begin{equation}
G = n K T_{C}^{(n-1)}.
\label{eq:ThermalCond}
\end{equation}
Both $K$ and $n$ change when a phononic filter is introduced, and therefore the flux and
thermal conductance can in principle change in different ways. In what follows, we shall
measure $K$ and $n$ directly for a variety of filters.

At first sight, it seems as if a suitable phononic filter might comprise alternating sections
of narrow and wide bars, but simulations indicate that is difficult to achieve large acoustic
impedance ratios, and the effect on power transmission is relatively small. More troublesome
is the fact that the dominant phonon wavelengths at 100 mK are of order 2 $\mu$m, or shorter,
and therefore optical lithography cannot be used easily to define steps that are highly
abrupt on a scale size of $\lambda / 4$, diminishing the effectiveness of the filter.

An alternative approach is to make the legs wider and introduce periodic patterns of holes,
thereby creating a truly phononic lattice \cite{PhononWaveBandgap,ComprehensiveTopology,HeatGuidingFocusing}. Such phononic crystals have been employed, for example, as support structures for micro-mechanical resonators, to reduce coupling loss due to elastic wave propagation to the substrate \cite{mohammadi2009high,hsu2011reducing,feng2014phononic}. Although this approach produces good
filter characteristics, the number of transmission channels available, prior to the filter
characteristic being applied, is high. Another way of thinking about this same problem is that
the phononic lattice comprises a large number of low-dimensional links, each of which transports
at least 4 modes. Thus the filter characteristic must compensate for the large increase in the
number of underlying modes simply to break even.

We have taken a different approach based on few-mode elastic interferometers and ring resonators:
An interferometer is formed by dividing a leg into two paths, one of which is longer than the
other. Simulations based on multimode travelling wave calculations \cite{DJThesis} indicate that
flux reductions of 25-75 \% are possible, depending on the number of interferometers used in
series. We have measured the thermal elastic attenuation length in SiN$_{\rm x}$ to be 20\,$\mu$m,
and given that this is much larger than a typical wavelength, individual filters behave in a phase
coherent way. Large series arrays of interferometers, however, become comparable to the
attenuation length, and so operate in the diffusive to ballistic transition. To
some extent, absorption isolates the effects of one interferometer on another. In other words,
locally the structure behaves as a phase-coherent few-element filter, but globally, the structure
conducts diffusively and behaves phase incoherently.

For the purposes of this paper, we define interferometers to be two-path elements that divide
and recombine the travelling waves gradually. Another option is a ring resonator design, in which case one benefits from the modal scattering that takes place at the junctions, as well
as the interferometric effects of the ring.

\subsection{Effective thermal response time}

In the results that follow we report direct measurements of thermal fluxes in multi-stage interferometers
and ring resonators patterned into the low-dimensional legs of SiN$_{\rm x}$ TESs. Furthermore, as an additional
indicator of reduction in thermal differential conductance, we also measure the time constants of phononic
devices.

\begin{figure}
\centering
\includegraphics[width=3.37in]{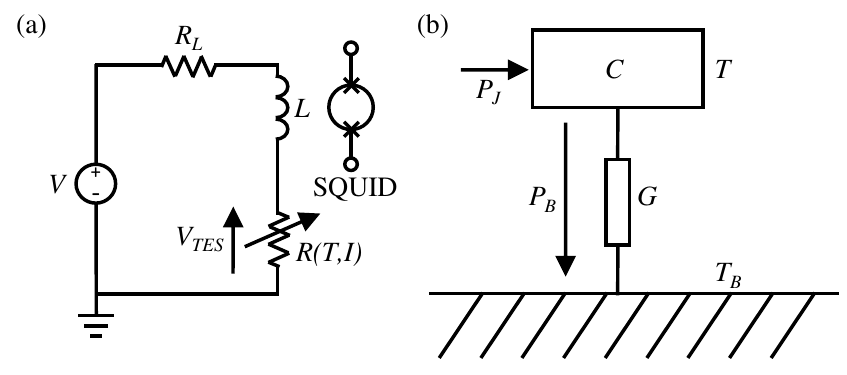}% Here is how to import EPS art
\caption{\label{fig:Circuits} (a): Th\'evenin equivalent representation of the TES bias circuit,
where the TES is shown as a variable resistor $R(T,I)$. $R_L$ is the internal resistance of the
voltage source, $V$. $R_L$ is the sum of a 1.45\,m$\Omega$ bias resistor and a $\approx 1$\,m$\Omega$
stray resistance. $L$ corresponds to the input inductance of the SQUID and any stray wiring inductance. $V_{TES}$ is the voltage across the TES.
(b): Thermal circuit showing the TES with heat capacity $C$ coupled to the heat bath via thermal
conductance $G$. $T$ and $T_B$ are the temperatures of the TES and the heat bath respectively .
$P_J$ represents the Joule power dissipated in the TES, and $P_B$ the thermal power flowing to the
heat bath.}
\end{figure}

The effective thermal time constant $\tau_{eff}$ of a TES may be determined from its response to a small
step in bias voltage. The induced current response may be derived from the coupled differential equations
that describe the TES electrical and thermal circuits. Figure \ref{fig:Circuits}(a) shows a Th\'evenin-equivalent
representation of a bias circuit connected to a TES having a current and temperature dependent resistance
$R(T,I)$, where the current is read out using an inductively coupled SQUID circuit. $R_L$ is the sum of the bias and stray
resistances, and $L$ represents the input inductance of the SQUID and any additional stray inductance due to wiring.
Figure \ref{fig:Circuits}(b) shows a representation of the thermal circuit, where to a first approximation, the
TES has a single heat capacity, $C$, coupled to the heat bath via support legs with thermal conductance, $G$.
The differential electrical and thermal equations are then
\begin{equation}
L \frac{\mathrm{d}I}{\mathrm{d}t} = V-IR_L-IR(T,I),
\end{equation}
and
\begin{equation}
C \frac{\mathrm{d}T}{\mathrm{d}t} = -P_B+P_J,
\end{equation}
respectively.
$T$ is the temperature of the central island, $P_B$ is the power flow to the heat bath, and
$P_J= I^{2} R(T,I)$ is the Joule power dissipated in the TES bilayer. Notice that we distinguish
between $T_{C}$, which is the critical temperature of the bilayer defined by some point on the
superconducting transition, and $T$, which is the temperature of the bilayer as the instantaneous
operating point moves up and down the transition.

It is standard practice in TES physics, to expand non-linear terms such as $P_B$, $P_J$ and $R(T,I)$
to first order in the small-signal limit around the steady state operating point $T_0$, $I_0$ and $R_0$,
giving \cite{IrwinChapter}
\begin{equation}
\frac{\mathrm{d}}{\mathrm{d}t}
\begin{pmatrix}
\delta I \\
\delta T
\end{pmatrix}
=-
\begin{pmatrix}
\frac{1}{\tau_{el}} & \frac{\alpha P_{J0}}{T_0I_0L} \\[0.7em]
-\frac{I_0R_0(2+\beta)}{C} & \frac{1}{\tau_I}
\end{pmatrix}
\begin{pmatrix}
\delta I \\
\delta T
\end{pmatrix}
+
\begin{pmatrix}
\frac{\delta V}{L} \\[0.5em]
0
\end{pmatrix},
\label{eq:Diff_Matrix}
\end{equation}
where $\delta I = I-I_0$, $\delta T = T-T_0$, $P_{J0} = I_0^2R_0$, and $\delta V$ represents a
small change in the applied bias voltage. The resistance-temperature and resistance-current sensitivities
are given by $\alpha = (\partial \ln R/\partial \ln T)_I$ and $\beta = (\partial \ln R/\partial \ln I)_T$
respectively. The time constants $\tau_{el} = L/(R_L+R_0(1+\beta))$ and $\tau_I = \tau/(1-P_{J0}\alpha/(GT_0))$
represent electrical and thermal time constants. The natural thermal time constant in the absence of
electrothermal feedback, $\alpha = 0$, and strictly $\beta = 0$, is given by $\tau = C/G$.

Adapting the approach of Lindeman\cite{LindemanThesis,IrwinChapter}, Eq. (\ref{eq:Diff_Matrix}) may be solved for
the specific case of a small step in bias voltage, $\delta V$ at $t=0$, subsequently maintained over the
course of a measurement, giving
\begin{equation}
\delta I = \frac{\delta V}{L}\frac{1}{\lambda_+-\lambda_-}(A_-\mathrm{e}^{-\lambda_-t}-A_+\mathrm{e}^{-\lambda_+t}+B),
\label{eq:RtMod}
\end{equation}
where $\lambda_\pm = \tau_\pm^{-1}$ are eigenvalues of the matrix in Eq. \ref{eq:Diff_Matrix},
$A_\pm = 1-\lambda_\pm^{-1}\tau_I^{-1}$ and $B=\tau_I^{-1}(\lambda_-^{-1}-\lambda_+^{-1})$.
For low inductance, $\tau_+ \ll \tau_-$, such that
\begin{align}
\tau_+ & \to \tau_{el} \\
\tau_- & \to \tau_{eff},
\end{align}
where $\tau_{eff}$ is the effective thermal time constant, given by
\begin{equation}
\tau_{eff}  = \tau\frac{1+\beta+R_L/R_0}{1+\beta+R_L/R_0+(1-R_L/R_0)P_{J0}\alpha/(GT_0)}
\end{equation}
\begin{equation}
\label{eq:tauEff_Approx}
\approx \frac{\tau}{1+\frac{\alpha}{n}(1-  T_B^n  / T_0^n )}.
\end{equation}
Equation \ref{eq:tauEff_Approx} is a simplified form following the assumptions that $\beta$ and
$R_{L}/R_{0}$ are small such that $R(T,I)\approx R(T)$, and the device is driven from a near-perfect
voltage source \cite{irwin1998thermal,IrwinThesis}. The empirical expression $P_{J0} = K(T_0^n-T_B^n)$
was used here, Eq. (\ref{eq:ThermalFlux}), which is standard in the TES community.

The time constant $\tau_{eff}$ governs the rate at which the current stabilises after a voltage step has been
applied, through the dominant exponential term in Eq. \ref{eq:RtMod}. This response time is significantly
shortened from its natural value $\tau = C/G$ due to negative electrothermal feedback when the TES is
voltage-biased in its transition, where $\alpha \gg 0$. Since $\tau_{eff}$ is approximately proportional
to $C/G$, a TES with reduced $G$ is expected to have a larger $\tau_{eff}$, which can be tested  experimentally
by fitting  Eq. \ref{eq:RtMod} to $\delta I(t)$. Measurements of $\tau_{eff}$ therefore provide an independent,
relative measure of the differential thermal conductances of devices, as distinct from the thermal fluxes,
assuming of course that the heat capacities of the devices are the same.

\section{Experiment}
\label{sec:Experiment}

\begin{figure}
\centering
\includegraphics[width=3.37in]{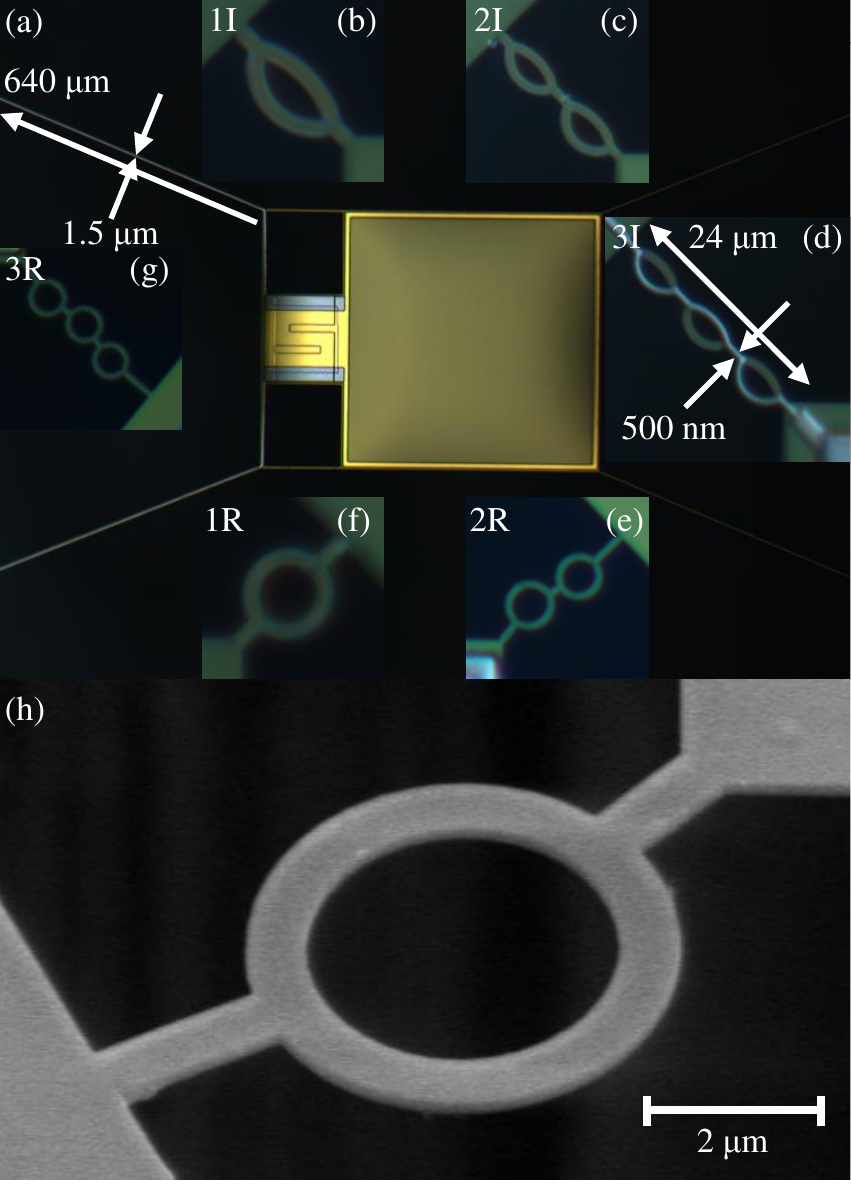}
\caption{\label{fig:TES_Images} (a) Optical microscopy of a typical ultra-low-noise TES with four long
diffusive SiN$_{\rm x}$ support legs and a far-infrared absorber. Insets (b)-(g) show representative
phononic legs of each type tested. These are labelled mI for an interferometer style, or mR for a ring
resonator style, where m is the number of filters in series. Inset (h) shows a triple interferometer
with Nb wiring on the SiN$_{\rm x}$. The cross section of each bar was 500 $\times$ 200\,nm for all
devices tested. Note the difference between the interferometers and ring resonators. (g) Scanning Electron Microscopy image 
of a 1R phononic leg, viewed obliquely.}
\end{figure}

Transition Edge Sensors having a variety of patterned phononic legs were fabricated on 200\,nm thick, low-stress,
amorphous SiN$_{\rm x}$ membranes.  Every TES had an identical $80 \times 80$\,$\mu$m
MoAu bilayer with 3 gold bars deposited on the upper surface, giving transition temperatures $T_C = 135 \pm 4$\,mK.
Phononic structures were classified according to the number of filters in series per leg, m, and the filter style.
Distinct filter styles were termed  either `interferometers' (mI), with pointed elliptical loops connected by
collinear microbridges, or `ring resonators' (mR), with typically angled connections intersecting circular rings:
Fig. \ref{fig:TES_Images}. The primary difference between the two styles lies in the way in which power is divided
and recombined upon entering and leaving a filter section: see later. A number of straight-leg control devices
(mIC) were also fabricated, with lengths equal to the direct end-to-end lengths of the interferometers mI.

In previous work we have always used optical lithography (OL) and reactive ion etching (RIE) to pattern
the the SiN$_{\rm x}$, followed by deep reactive ion etching (DRIE) to release the membrane from its supporting
Si substrate\cite{glowacka2012fabrication}. Through this method we have been able to fabricate narrow legs, down to $W=$ 700\,nm, with
a high degree of reliability and reproducibility. This method was also used to fabricate our previous few-mode
ballistic devices, and we have successfully produced prototype interferometers using OL. The devices reported
in this paper, however, used EBL and RIE to pattern the membranes. This required the development of direct-write
EBL processing to pattern the legs and define the sputtered Nb bias leads. These new techniques then had to
be combined with conventional OL to fabricate the main body of the TES. Using this hybrid method, we have been
able to fabricate interferometers and ring resonators having leg cross sections of only 300 $\times$ 200\,nm,
which ensures that only 4 elastic modes are excited in each bar of a structure for temperatures below 100\,mK.

Figure \ref{fig:TES_Images}(a) shows one of our traditional ultra-low-noise devices having long, straight
legs ($W =$ 1.5\,$\mu$m, $L =$ 640\,$\mu$m). The MoAu bilayer, and bars, can be seen as a small gold-coloured
square with lateral bars, and the $\beta$-phase Ta FIR absorber as a large gold-rimmed square. Around the outside of
Fig. \ref{fig:TES_Images}(a), (b - g), we show a number of the phononic filters fabricated.  These comprised
single,  double and triple interferometers and ring resonators, and all of the features had cross sections
of 500 $\times$ 200\,nm. Table \ref{tab:TES_Props} lists the devices tested, with the path difference $\Delta L$
engineered between the arms of the filters in each phononic leg. Figure \ref{fig:TES_Images}(d) shows the
Nb wiring, for bias and readout, on one of the 3-element interferometers, with an alignment tolerance of
50\,nm. Nb is significantly less stiff than SiN$_{\rm x}$ and therefore does not influence the elastic modes
of the structure even though its thickness is comparable with that of the SiN$_{\rm x}$. The superconducting
wiring also contributes negligible electronic heat conduction because the quasiparticle density is exceedingly
small at low temperatures. Figure \ref{fig:TES_Images}(h) shows a Scanning Electron Micrograph of a 1R phononic leg, viewed obliquely.

Figure \ref{fig:TES_Images} illustrates that it is possible to fabricate few-mode multi-element interferometers,
with 500\,nm wide features, outstanding definition, and well-aligned Nb wiring. It is remarkable that these
tiny patterned legs are perfectly able to support the main body of the TES, and can be fabricated with
high yield, which was due in part to our ability to control film stresses in the main body of the device.
As will be seen later, it is also notable that these devices performed perfectly well as TESs, with
no evidence of anomalous behaviour, such as weak links or additional stray resistance where the Nb
leads meandered over the arms of the interferometers. As will be seen later, the thermal properties of these
tiny structures were fully consistent with few-mode elastic behaviour even though they were supporting the
relatively large central island of the TES. This occurs because the bulk elastic constants are relatively
insensitive to static strain, and furthermore the dispersion relationships are insensitive to the bulk
elastic constants.

\begin{table*}
\caption{\label{tab:TES_Props} Characteristics of the devices tested. $\Delta L$ is the path length
difference between the arms of each filter, $n$ the power-flow temperature exponent, $K$ the
power-flow scaling factor, $T_C$ the critical temperature, $\epsilon$ the power flow normalised
to the few-mode ballistic case, $G$ the thermal conductance from the TES island to the heat bath,
$N_{eff}$ the effective number of elastic modes per leg, and $\tau_{eff}$ the effective thermal
time constant. Devices types are labelled mI$_x$ for an interferometer, mR$_x$ for a ring resonator,
and mIC$_x$ for a straight-leg control, where m is the number of filters in series, and x distinguishes
multiple devices of the same type.}
\begin{ruledtabular}
\begin{tabular}{clccccccc}
Device & $\Delta L$ ($\mu m$) & $n$ & $K$ (pW/K$^n$) & $T_C$ (mK) & $\epsilon$ & $G$ (pW/K)  & $N_{eff}$ & $\tau_{eff}$ (ms)\\
\hline
1IC$_a$ & - & 2.70 & 22.2 & 141.1 & 0.66 & 2.16 & 3.39 & 0.54\\
1IC$_b$ & - & 2.53 & 15.4 & 134.5 & 0.62 & 1.81 & 3.13 & 0.64\\
2IC$_a$ & - & 2.43 & 14.2 & 137.4 & 0.69 & 2.01 & 3.47 & 0.53\\
2IC$_b$ & - & 2.37 & 12.3 & 133.3 & 0.67 & 1.86 & 3.36 & 0.64\\
3IC$_a$ & - & 2.28 & 7.5 & 138.3  & 0.48 & 1.36 & 2.42 & 0.85\\
1I$_a$ & 1 & 2.74 & 20.1 & 141.4  & 0.55 & 1.84 & 2.83 & 0.60\\
1I$_b$ & 1.5 & 2.59 & 13.7 & 131.7  & 0.49 & 1.41 & 2.44 & 0.78\\
1R$_a$ & 0 & 2.69 & 15.2 & 137.2   & 0.45  & 1.42 & 2.30 & 1.20\\
2I$_a$ & 1, 1.5 & 2.62 & 11.8 & 127.8 & 0.40 & 1.10 & 1.98 & 0.90\\
2I$_b$ & 1, 1.75 & 2.67 & 13.3 & 134.5 & 0.41 & 1.25 & 2.07 & 0.73\\
2R$_a$ & 1, 1.75 & 2.60 & 8.1 & 127.4  & 0.28  & 0.78 & 1.39 & 1.27\\
3I$_a$ & 1, 1.5, 2 & 2.47 & 6.5 & 136.7 & 0.29  & 0.86 & 1.48 & 1.04\\
3I$_b$ & 1, 1.25, 1.75 & 2.48 & 6.7 & 133.2  & 0.30 & 0.85 & 1.49 & 1.52\\
3R$_a$ & 1, 1.25, 1.75 & 2.44 & 4.0 & 128.5 & 0.19  & 0.51 & 0.94 & 1.88\\
3R$_b$ & 1, 1.3, 1.6 & 2.51 & 4.4 & 129.3 & 0.19  & 0.51 & 0.92 & 1.28\\
\end{tabular}
\end{ruledtabular}
\end{table*}

Each TES was voltage-biased with a low impedance source ($\approx$ 1.5\,m$\Omega$) and read out
using a two-stage SQUID amplifier as a low-noise current-to-voltage converter. The TES and
SQUID chips were mounted in an optically blackened light-tight box and cooled to a base
temperature of 68\,mK in an adiabatic demagnetisation refrigerator (ADR). The bath temperature
of the TES chip was taken to be that of the copper housing, held constant to within 200\,$\mu$K
by means of the residual current in the ADR magnet. Current and voltage offsets and stray
resistances were identified and compensated for in data processing. The TES current response to
a step in voltage was obtained by biasing the TES in its transition and superimposing a square wave
on the bias input, with small amplitude compared to the voltage width of the transition. Current
response was averaged over multiple leading-edge voltage steps.

\section{Results and Discussion}
\label{sec:Results_Discussion}

A TES voltage-biased within its transition self-regulates its temperature due to negative
electrothermal feedback. In the steady state, $dT / dt= 0$, the net power flow from the island
to the heat bath is equal to the Joule power dissipated in the bilayer. The power flow
is therefore given by $P_{B} = P_{J} = IV_{TES}$, allowing $P_{B}$ to be
obtained from a series of  $I$-$V_{TES}$ curves taken over a range of bath temperatures.
If the electrothermal feedback is strong, the TES temperature is essentially constant within the transition at $T_C$, allowing $K$ and
$n$ to be found.

\begin{figure}
\centering
\includegraphics[width=3.37in]{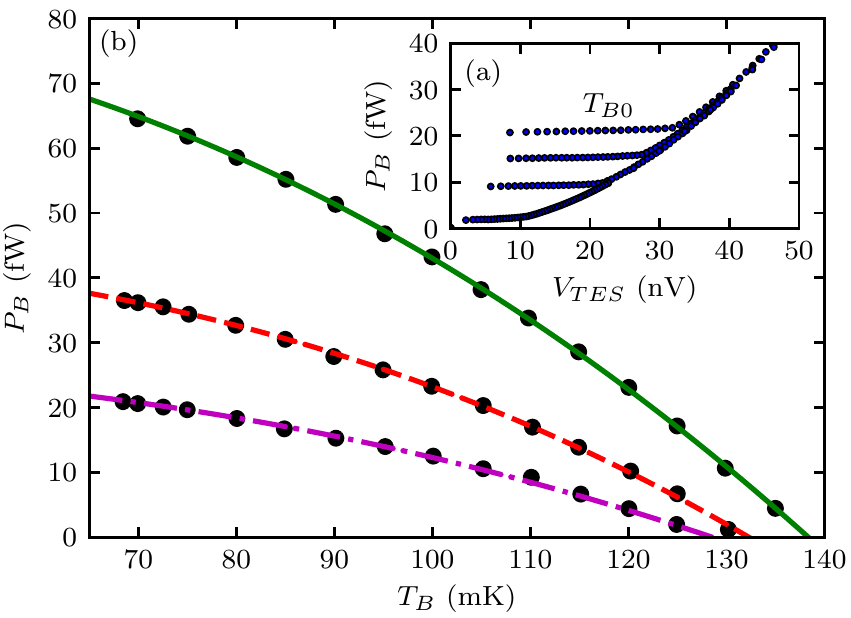}
\caption{\label{fig:PT_Inset_PV}
(a) Power against voltage across the TES, $V_{TES}$, for device 3R$_b$, for a subset of $T_B$
from $T_{B0}=67.5$\,mK to 125\,mK from top to bottom. (b) Net power flow from the TES island to
the heat bath, $P_{B}$, against bath temperature, $T_B$, for devices 3IC (green, solid), 3I$_b$
(red, dashed) and 3R$_b$ (magenta, compound dashed). Points show measured data, and lines
indicate model fits according to Eq. (\ref{eq:ThermalFlux}).
 }
\end{figure}

Figure \ref{fig:PT_Inset_PV}(a) shows thermal power against the voltage across the TES, $V_{TES}$,
for device 3R$_b$, for a set of bath temperatures, $T_{B}$. The topmost curve corresponds to the
lowest bath temperature used, $T_{B0}$. The power is essentially constant across the voltage range
for which the bilayer is in its transition, indicating the presence of strong electrothermal feedback.
This plot is representative of all devices tested, and demonstrates that the presence of the
phononic filters in the legs (4 for each device), and associated Nb wiring layer, does not introduce
artifacts into the operation of the device.

The power averaged over the transition region is shown in Fig. \ref{fig:PT_Inset_PV}(b)
as a function of $T_B$ for devices 3IC (green), 3I$_b$ (red), and 3R$_b$ (magenta).
Figure \ref{fig:PT_Inset_PV}(b) displays a clear reduction in the power transmitted through the
triple phononic filters relative to the straight leg control, with further improved attenuation
for the ring resonator 3R$_b$ over the interferometer 3I$_b$. This behaviour is reproduced
almost identically in devices 3I$_a$ and 3R$_a$, For all of the devices tested,
Eq. (\ref{eq:ThermalFlux}) was fitted to data of this kind under the assumption that the
temperature of the TES was maintained constant at nearly $T_C$, which is true for sufficiently sharp
transitions. $K$, $n$ and $T_C$ were free parameters in the fitting process, with $T_C$ corresponding to the
intercept of the curve on the $P_{B}=0$ axis.

For each device, the measured power at the lowest bath temperature, $T_{B0} = 68 \pm 1$\,mK, was used
to calculate the normalised power per leg, $\epsilon$, according to Eq. \ref{eq:Epsilon}. In this way, the
measured fluxes were normalised to the theoretical ballistic power for a device with straight legs with
termination temperatures $T_{B0}$ and $T_C$. The total thermal conductance $G$ of the support structure
was determined from Eq. \ref{eq:ThermalCond}. Table \ref{tab:TES_Props} lists measured values of
$n$, $K$, $T_C$, $\epsilon$, $G$, and $N_{eff}$.

\begin{figure}
\centering
\includegraphics[width=3.37in]{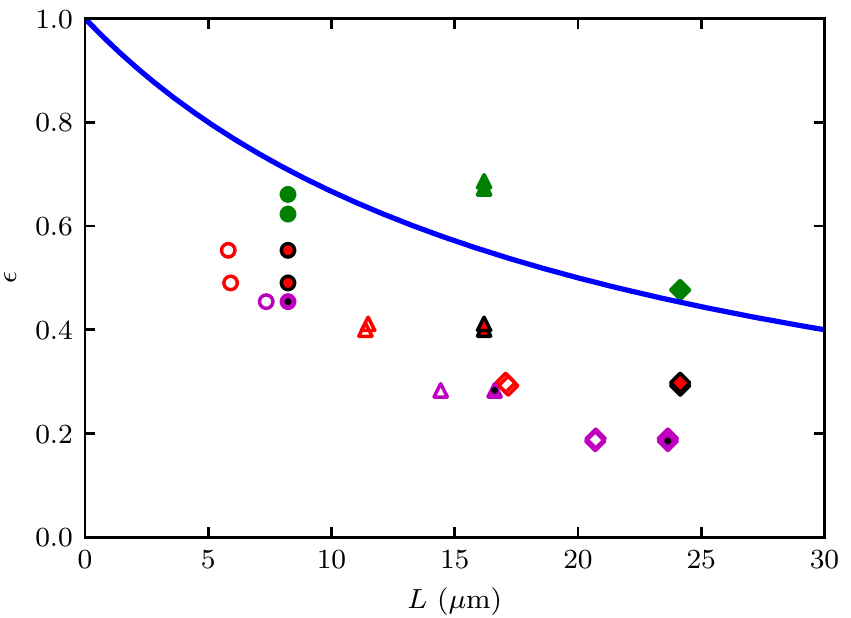}% Here is how to import EPS art
\caption{\label{fig:Epsilon_L}Net power flow through each leg, $\epsilon$, against leg length, $L$, for
the lowest bath temperature measured, $T_{B0}$, normalised to the theoretical ballistic power for a
straight leg, given $T_{B0}$ and $T_C$. Straight leg control, interferometer and ring
resonator structures are shown in green, red with border and magenta with central dot respectively.
Circles, triangles and diamonds represent legs with one, two and three filters, with control devices
matching their corresponding phononic designs. Filled markers are plotted with respect to the direct
end-to-end leg length. For phononic legs, open markers show the same $\epsilon$ against the equivalent
length of a straight leg that would give the same thermal conductance in a purely diffusive model.
The blue line shows $\epsilon_{r}$ for diffusive phonon transport according to Eq. (\ref{eq:Epsilon_L_Diff})
with an acoustic attenuation length of 20\,$\mu$m.}
\end{figure}

Figure \ref{fig:Epsilon_L} shows the normalised flux $\epsilon$ against leg length $L$ for all of
the devices tested. Also shown for comparison (solid blue line) is an analytical model for heat
transport in the diffusive to ballistic regime:
\begin{equation}
\epsilon_{r} =  (1+L/L_a)^{-1}.
\label{eq:Epsilon_L_Diff}
\end{equation}
In previous work\cite{ThermalAttenuation2017}, we determined the acoustic attenuation length,
$L_a$, to be 20\,$\mu$m in SiN$_{\rm x}$ at low temperatures. This was achieved by fitting
Eq. (\ref{eq:Epsilon_L_Diff}) to data from a set of straight leg devices having lengths,
1 - 490\,$\mu$m, which span the diffusive to ballistic transition. An $\epsilon$ of less than
unity indicates that a filter has a transmission factor lower than the ballistic case, and
an $\epsilon$ of less than $\epsilon_{r}$ indicates that a filter has a transmission factor
lower than its straight-legged counterpart, where some diffusive scattering is present.

In order to compare the flux of a phononic filter with a straight-legged device, it is
necessary to assign an equivalent length to the filter, and this can be done in a variety
of ways. In Fig. \ref{fig:Epsilon_L}, filled markers show the normalised flux as a
function of the overall end-to-end length of each phononic leg, equal to the length of the
corresponding straight reference legs spanning the same gap. It could be argued, however,
that the actual length of the path travelled should be used.  For a purely diffusive process,
where $G \propto 1/L$, the greater path length of a curved leg would reduce $G$ relative to a straight
leg device with the same end-to-end length irrespective of any coherent destructive process.
This should be taken into account, but it is still not clear which path along a multistage
filter should be used.

For a fully diffusive process, it is possible to define an equivalent length, $L_{eq}$,
based on the notion of thermal conductances in parallel. In a single interferometer for example,
$L_{eq} = L_{l1}+(1/L_{a1}+1/L_{a2})^{-1}+L_{l2}$, where $L_{l1}$ and $L_{l2}$ correspond
to the straight linking sections and $L_{a1}$ and $L_{a2}$ to the lengths of the different
paths around the filter. $L_{eq}$ is therefore a single equivalent length giving the
same $G$ as a chain of series and parallel conductances representing a phononic structure, for $G \propto 1/L$. 
This constitutes a more appropriate definition of length in
Eq. \ref{eq:Epsilon_L_Diff} for phononic structures, because it does not mistakenly imply
that a reduction in $\epsilon$ due to the longer path length of the interferometer is
necessarily due to coherent interference. The open markers in Fig. \ref{fig:Epsilon_L} show
$\epsilon$ against $L_{eq}$ for all phononic devices. All of the open markers are to the
left of the solid markers because the parallel arms reduce the effective length.

The phononic legs show a clear reduction in transmitted power relative both to
their corresponding straight leg control devices and the diffusive attenuation expected
from Eq. \ref{eq:Epsilon_L_Diff}. This presents strong evidence that micromachined phononic
filters can be used to reduce thermal flux. Moreover, the reductions achieved are comparable
with those predicted previously\cite{DJThesis}. A maximum flux reduction to 19 \% of the
ballistic limit is achieved for the 3R devices, corresponding to 38 \% of the flux in fully
diffusive devices. From Eq. \ref{eq:Epsilon_L_Diff}, a leg length of 87\,$\mu$m would be
necessary to achieve this attenuation in the absence of the phononic filter, a more than
threefold increase from the 24\,$\mu$m end-to-end length actually used. The expected
monotonic decrease in $\epsilon$ with number of filters per leg is also observed within
both the interferometer and ring resonator groups \cite{DJThesis}.

Within pairs of devices of the same type, the greatest difference in $\epsilon$ for different
filter path lengths, $\Delta L$, is observed between 1I$_a$ and 1I$_b$, with the larger
$\Delta L$ giving the lower transmission. For devices of type 2I, 3I and 3R, variations in
$\Delta L \leq 0.25$\,$\mu$m of the second and third filter stages show negligible effect on $\epsilon$.
This insensitivity is reasonable since a change in differential path length, $\Delta L$, shifts
the fringe of the filter, in frequency space, relative to the wide band blackbody spectrum,
changing the transmitted flux very little. In the next phase of the work, we will carry
out detailed simulations of precise designs in order to understand the degree to which modelling
can be used to predict and optimise behaviour.

Figure \ref{fig:Epsilon_L} shows that ring resonators perform significantly better than their
interferometer counterparts, for single, double and triple designs, including the case where
$\Delta L$ is the same for the two types. The origin of this improvement is likely to be
due to the way in which the principle modes scatter at the junctions. An elastic wave reaching
an interferometric filter may maintain its symmetry as it divides between the two arms, whereas
in a ring resonator, the incoming wave encounters a perpendicular bar and mode conversion
takes place; for example from a torsional wave to two out of phase flexures. Similarly, ring
resonators allow waves to propagate multiple times around the ring, possibly contributing to
Fabry-P\'{e}rot-like enhanced attenuation.

At the lowest bath temperatures, $T_B\approx68$ mK, used in this experiment, the effective number 
of modes transporting heat in a purely ballistic leg is approximately 5, which is very close to the 
quantised limit of 4. For $\epsilon_{r} = 0.5$, from Eq. \ref{eq:Epsilon_L_Diff} for $L=L_{eq}=20.7$ 
in the case of the 3-element ring resonator, the calculated effective number of modes is then 2.5.  
However, Table \ref{tab:TES_Props} shows that for both 3R devices, the effective number of modes 
is 0.92, from the measured power. Thus the phononic structures have significantly reduced the effective 
number of modes through frequency-domain filtering.

\begin{figure}
\centering
\includegraphics[width=3.37in]{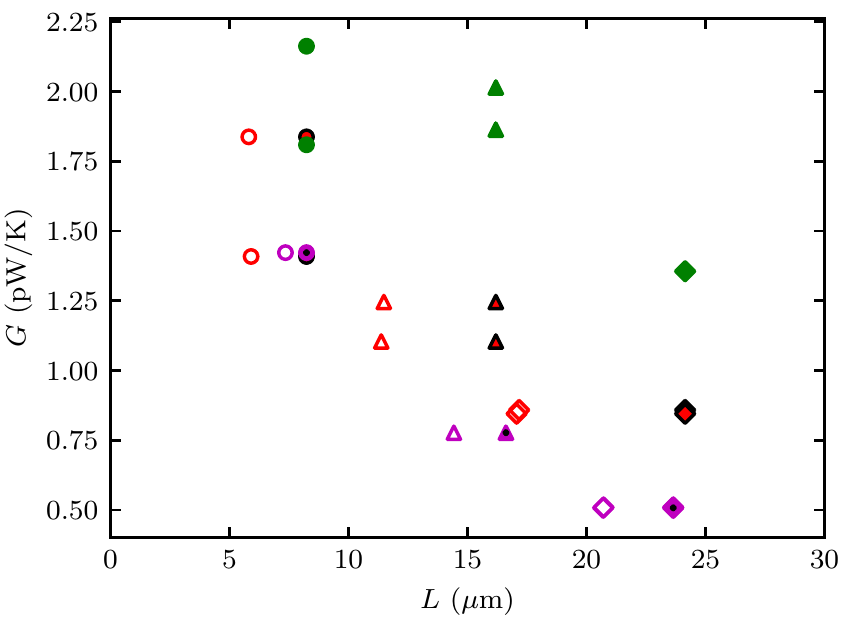}
\caption{\label{fig:G_L} Conductance, $G$, against leg length, $L$, for the lowest bath temperature
measured, $T_{B0}$, Straight leg control, interferometer and ring resonator structures are shown in
green, red with border and magenta with central dot respectively. Circles, triangles and diamonds
represent legs with one, two and three filters, with control devices matching their corresponding
phononic designs. Filled markers are plotted with respect to the direct end-to-end leg length. For
phononic legs, open markers show the same $\epsilon$ against the equivalent length of a straight
leg that would give the same thermal conductance in a purely diffusive model.}
\end{figure}

Figure \ref{fig:G_L} plots the differential conductance $G$ against leg length. Indeed
the small-signal behaviour of a TES depends on $G$ rather than on the absolute value of flux. The
trends in conductance are essentially the same as those in flux, with minor differences due to the effect 
on $G$ of variations in $T_C$ between devices. The 2- and 3- stage ring resonators significantly
reduce the conductance below the ballistic value of about 2.2 pW\,K$^{-1}$.  In the case of phononic
filters, Table \ref{tab:TES_Props} shows that the reduction in $\epsilon$ and $G$ is associated with a reduction
in $K$, with $n$ staying almost constant. In the ballistic case, we find $n \approx 2.5$,\cite{DJBallistic}
which is  slightly above the single-mode value $n=2$. In the case of phononic filters, it seems that
$n \approx 2.5$ also. This is very different to the case of long, narrow diffusive legs where $n$ takes
on values of unity and below, which we have always regarded as being an indicator of the effects of TLS loss
in the disordered SiN$_{\rm x}$ \cite{ThermalAttenuation2017}.

The values of $G$ achieved with few-mode ballistic and phononic legs are already highly suitable for
many applications, but in particular it should be noted that if we were to use a 3-stage ring resonator
with $T_{B} =$ 50\,mK and $T_{C} = $ 100\,mK, then $G =$ 0.3\,pW\,K$^{-1}$, and we would be close to the
requirement $G= 0.2$\,pW\,K$^{-1}$ for the ultra-low-noise TESs needed for SPICA. Now, however, the
legs would only be of order 25\,$\mu$m long, rather than the 600-700\,$\mu$m long legs currently used.

Of particular note is the remarkable consistency in $\epsilon$ and $G$ between different devices of
similar design. In fact some of the points on Fig. \ref{fig:Epsilon_L} and Fig. \ref{fig:G_L} are
difficult to distinguish. In conventional TESs having narrow, straight legs, hundreds of microns long,
different research teams see conductance variations of $\pm15\%$ or higher between notionally
identical devices, even from the same wafer. This variation is attributed to phonon localisation,
where elastic waves are reflected by impedance discontinuities due to disorder in the dielectric,
creating resonant cells that exaggerate variations in elastic properties\cite{ThermalAttenuation2017}.
The reproducibility seen in Fig. \ref{fig:Epsilon_L} strongly suggests that phononic filters
are capable of producing highly uniform arrays, eliminating the troublesome effects of localisation
seen in conventional devices.

\begin{figure}
\centering
\includegraphics[width=3.37in]{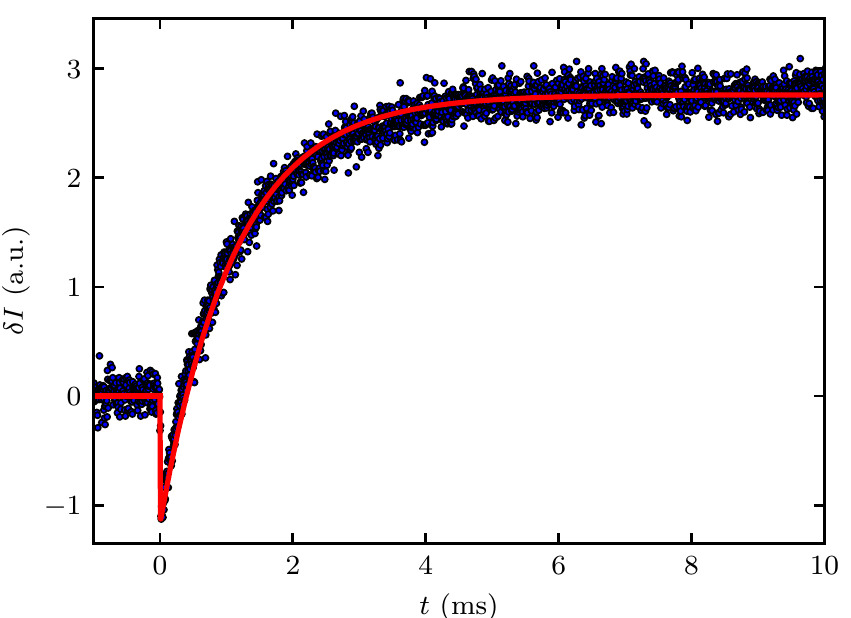}% Here is how to import EPS art
\caption{\label{fig:Rt_Fit} TES current response, $\delta I$, to a small step increase in voltage at
time $t=0$, in arbitrary units for device 3R$_b$ at 90.3\,mV, 28\% of the TES normal resistance. The
red line shows $\delta I$ according to the model given by Eq. \ref{eq:RtMod}: $\tau_{eff} = 1.12$\,ms,
$\tau_{el} = 2.16$\,$\mu$s, $\tau_{I} = -0.48$\,ms.}
\end{figure}

As an independent indicator of the reduction in $G$, we measured the effective thermal time constants
of the phononic TESs. Because all of the TESs were identical, apart from the different leg designs, we
would expect the time constants to follow $G$ in the appropriate way. Figure \ref{fig:Rt_Fit}  shows
the change in current, $\delta I$, in response to a small step in bias voltage at $t=0$ for
device 3R$_b$ at 90.3\,mV, corresponding to a point on the transition where the resistance of the
bilayer was 28 \% of its normal value. The first dip on the leading edge is due to the electrical
response of the TES, and its bias circuit, whereas the slowly rising trailing edge is due to the
electrothermal relaxation.

$\delta I(t)$ given by Eq. (\ref{eq:RtMod}) was fitted to the measured data with a scaling pre-factor to
give $\alpha = 551$ and $\beta = 0.85$, corresponding to $\tau_{eff} = 1.12$\,ms, $\tau_{el} = 2.16$\,$\mu$s
and $\tau_{I} = -0.48$\,ms, for this particular bias point. The steady state values of $I_0$, $R_0$ and $P_{J0}$ were taken from $I$-$V_{TES}$ measurements, $T_0=T_C$ was assumed, $L$ was derived from impedance measurements with the bilayer
in its fully superconducting state, and $C=41.8$\,fJ\,$K^{-1}$ was calculated using the volumes and specific heats
of the various materials used. We found that although the values of $\alpha$ obtained scale with the value
of $C$ assumed, the fitted values of $\tau_{eff}$ obtained do not change; in other words, there
is a linear degeneracy between the $\alpha$ and $C$, but the same value of $\tau_{eff}$ always results.
Figure \ref{fig:Rt_Fit} is typical of the data taken, and the fit is in good agreement with the model
despite using only a single heat capacity $C$.

\begin{figure}
\centering
\includegraphics[width=3.37in]{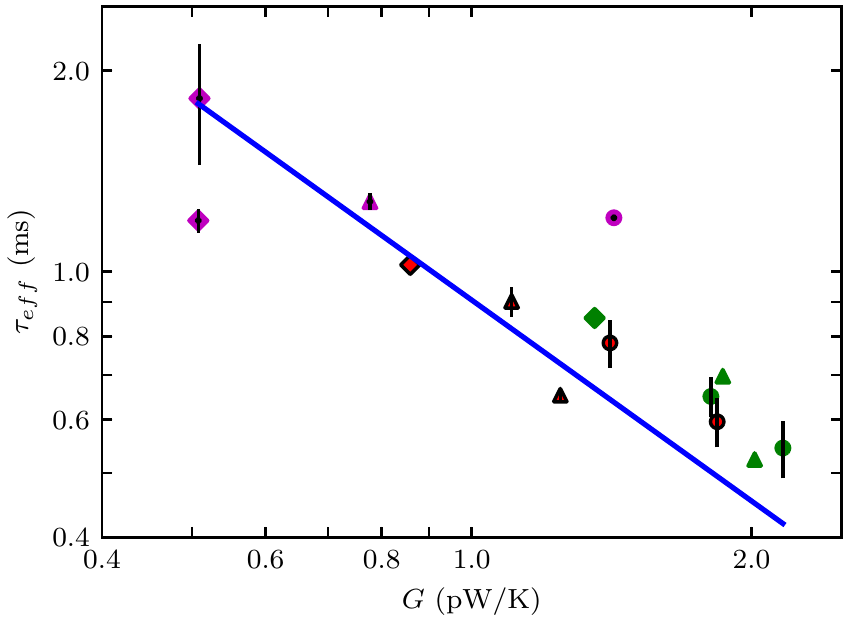}% Here is how to import EPS art
\caption{\label{fig:TauEff_G} Effective thermal time constant, $\tau_{eff}$, against thermal conductance $G$
for all devices. As in Fig. \ref{fig:Epsilon_L}, straight leg control, interferometer and ring resonator
structures are shown in green, red with border and magenta with central dot respectively. Circles, triangles
and diamonds represent legs with one, two and three filters, with control devices matching their corresponding
phononic designs. Error bars correspond to standard error in the mean of multiple measurements of $\tau_{eff}$
at different bias points. The blue line shows the fitted model $\tau_{eff}\propto1/G$.}
\end{figure}

Figure \ref{fig:TauEff_G} plots $\tau_{eff}$ against $G$ for all of the devices tested. The error bars
correspond to the standard error in the plotted mean of multiple measurements of  $\tau_{eff}$ for different bias
points across the transition, where applicable for each device. $\tau_{eff}$ is expected to be approximately
inversely proportional to $G$ from Eq. (\ref{eq:tauEff_Approx}), and so the simple model $\tau_{eff}\propto  1/
\/G$ is shown as a blue line on Fig. \ref{fig:TauEff_G}. The general agreement offers additional evidence that
the phononic filters reduced the differential conductance of the legs in the expected way. There are various
reasons why, however, this proportionality may not be exact. The heat capacity will increase slightly as bias
voltage is reduced through the transition as the MoAu bilayer becomes increasingly superconducting.
Additionally, unknown sources of heat capacity may exist, for example due to residual SiO$_2$ used as an etch
stop in fabrication. The steady state values $I_0$, $R_0$ and $P_{J0}$ vary with bias point, although the impact
on $\tau_{eff}$ between measurements on the same device is typically small. Notwithstanding these
considerations, Fig. \ref{fig:TauEff_G} shows an overall decrease in $\tau_{eff}$ with increasing $G$, which
implies that the interferometrically reduced values of $G$ derived through the measured values of $K$ and $n$
are true differential conductances.

\section{Conclusions}

We have successfully manufactured a range of superconducting transition edge sensors having few mode, phononic
thermal isolation in the legs. By using electron beam lithography we were able to pattern interferometers
and ring resonators into legs having cross-sectional dimensions of only 500 $\times$ 200\,nm. At temperatures
of around 100\,mK each leg effectively transports heat in just 5 elastic modes, which is close to the quantised limit of 4.
The phononic filters then reduced the thermal flux and conductance further. Nb bias leads were patterned on
the filters to an alignment tolerance of better than 50\,nm. The manufacturing process proved to be highly
reliable, giving robust devices with high dimensional definition.

Significant reductions in thermal flux and thermal conductance were recorded, with the ring resonators giving
the highest rejection ratios. No artifacts were seen in behaviour, making the devices suitable for many
applications. The device-to-device variation in thermal conductance of notionally identical devices was
exceedingly small, and well below the $\pm$ 15 \% frequently seen in conventional long-legged designs. It
should also be noted that the temperature exponent $n$ stayed at its near ballistic value of 2.5, in contrast to the case of long narrow legs, where diffusive transport due to two level systems reduces the
exponent to typically 1, and below.

A key advantage of phononic filters is that it is possible to approach the very lowest $G$s seen with long
($>$ 700\,$\mu$m) diffusive legs, but using significantly shorter structures. The attenuation length of the low-order
modes in SiN$_{\rm x}$ is 20\,$\mu$m, and the ring resonators are typically 5\,$\mu$m in diameter, and therefore by
placing a large number of ring resonators in series, dividing a diffusive leg into phase-coherent phononic cells,
one would expect to be able to realise thermal conductances significantly smaller than anything achieved to date.
Also, one would expect to be able to create ultra-low-noise TESs using crystalline ballistic Si membranes, which
would have many advantages.

The next stage in our work will be to carry out a range of scattered travelling wave simulations in diffusive
structures in an attempt to identify optimised filters with even higher levels of attenuation. We have already
developed a modelling technique for patterned phononic structures operating in the ballistic to diffusive regime,
and this will be reported in an upcoming paper.

\begin{acknowledgments}

The authors are grateful to Science and Technology Facilities Council for funding this work.
Emily Williams is grateful for a PhD studentship from the NanoDTC, Cambridge, EP/L015978/1.

\end{acknowledgments}

%\nocite{*}
\bibliography{References}% Produces the bibliography via BibTeX.

\end{document}